\documentclass[12pt,preprint,psfig]{aastex}

\usepackage{emulateapj5}

\def\ltsima{$\; \buildrel < \over \sim \;$}

\def\lsim{\lower.5ex\hbox{\ltsima}}
\def\gtsima{$\; \buildrel > \over \sim \;$}
\def\gsim{\lower.5ex\hbox{\gtsima}}
\begin{document}
\title{Undetected Sources Allow Transmission of the Ly$\alpha$ Line From
Galaxies Prior to Reionization}

\author{J. Stuart B. Wyithe\altaffilmark{1} and Abraham
Loeb\altaffilmark{2}}

\email{swyithe@isis.ph.unimelb.edu.au; aloeb@cfa.harvard.edu}

\altaffiltext{1}{University of Melbourne, Parkville, Victoria, Australia}

\altaffiltext{2}{Harvard-Smithsonian Center for Astrophysics, 60 Garden
St., Cambridge, MA 02138}

\begin{abstract}
\noindent 
The discovery of Ly$\alpha$ emission from galaxies at redshifts beyond
$z\sim6.5$ should not be naively interpreted as implying that the
intergalactic medium (IGM) had been reionized at higher redshifts.  We show
that a cluster of faint undetected sources around each observed galaxy
generates an HII region sufficiently large to allow transmission of the
galaxy's Ly$\alpha$ line prior to reionization.  We also show that quasars
may contribute a significant fraction of the ionizing photons to HII
regions around galaxies with a velocity dispersion larger than
$\sim100~{\rm km~s^{-1}}$.  These contributing quasars are not usually seen
due to the small fraction of time they spend in a luminous phase.

\end{abstract}

\keywords{cosmology: theory - galaxies: formation}

\section{Introduction}

The identification of a Gunn-Peterson trough in the spectra of the most
distant quasars at redshifts of $z\sim 6.3$--$6.4$ (Fan et al.~2004) hints
that the reionization of cosmic hydrogen was completed only around that
time, about a billion years after the big bang (White et al. 2003; Wyithe
\& Loeb~2004).  On the other hand, the recent discovery of Ly$\alpha$
emission from galaxies beyond a redshift of $z\sim6.5$ (Hu et al., 2002;
Kodaira et al., 2003; Rhoads et al. 2004; Kurk et al. 2004; Stern et
al. 2004) was naively interpreted as an indication that the surrounding
intergalactic medium (IGM) was highly ionized, in apparent conflict with
the quasar data. Although a galaxy is often surrounded by an HII region of
ionized hydrogen (Haiman~2002; Santos~2004), its Ly$\alpha$ line could be
heavily suppressed by the red damping wing of the resonant Ly$\alpha$
absorption in the surrounding IGM (Loeb, Barkana \& Hernquist~2004; Cen,
Haiman \& Messinger~2004; Ricotti et al.~2004). Clustering of sources may
result in HII regions that are larger than expected for a single galaxy,
rendering the optical depth from the red damping wing of the IGM irrelevant
(Gnedin \& Prada~2004; Furlanetto, Hernquist \& Zaldarriaga~2004).  Here we
show that undetected sources such as short-lived quasars or a cluster of
faint galaxies, are expected to dominate the growth of a galaxy's HII
region and allow transmission of its Ly$\alpha$ photons prior to
reionization.

\section{The Size of Ionized Regions}
The average mean-free-path for an ionizing photon in a fully neutral
IGM is much smaller than the size of a cosmological HII region. By
balancing the number of ionizing photons and neutral atoms we can
therefore estimate the characteristic physical size of a spherical HII
region surrounding a galaxy with a dark matter halo of mass $M_{\rm
  halo}$ (Loeb et al.~2004)
\begin{eqnarray}
\label{Dstar}
\nonumber
\nonumber D_{\star} &=& 0.75 \left(\frac{x_{\rm HI}}{0.1}\right)^{-1/3}
\left(\frac{M_{\rm
halo}}{10^{10}M_\odot}\right)^{1/3}\left(\frac{f_{\star}f_{\rm esc}/N_{\rm
reion}}{0.003}\right)^{1/3}\\
&\times&\left(\frac{N_{\star}}{4000}\right)^{1/3}
\left(\frac{1+z}{7.5}\right)^{-1}\mbox{Mpc},
\end{eqnarray}
where $f_{\star}$ is the star formation efficiency and $f_{\rm esc}$ is
the fraction of ionizing photons that escape from the galaxy into the
surrounding IGM of neutral hydrogen fraction $x_{\rm HI}$.
Here we have assumed that the number of ionizing photons per baryon
incorporated into (Pop-II) stars is $N_{\star}\sim4000$ (Bromm, Kudritzki \& Loeb~2001), and
that $N_{\rm reion}$ photons are required per baryon for the reionization
of the HII region.  The value of $N_{\rm reion}$ depends on the number of
recombinations and hence on the clumpiness of the IGM.
The halo mass $M_{\rm halo}$ is a function of its velocity dispersion
$\sigma$ [see Eqs.~22--25 in Barkana \& Loeb~(2001)], which we set equal
to $v_{\rm c}/\sqrt{2}$, where $v_c$ is the halo circular velocity.

Similarly, the physical size of the HII region generated by an accreting
supermassive black-hole of mass $M_{\rm bh}$ is (White et al., 2003)
\begin{eqnarray}
\label{Dquasar}
\nonumber
D_{\rm q} &=& 0.43 \left(\frac{x_{HI}}{0.1}\right)^{-1/3}
\left(\frac{M_{\rm bh}}{10^5M_\odot}\right)^{1/3} \left(\frac{{\dot N}_{\rm
\gamma,5}/N_{\rm reion}}{3\times10^{53}\mbox{s}^{-1}}\right)^{1/3}\\
&\times&
\left(\frac{f_{\rm dyn}}{0.035}\right)^{1/3}
\left(\frac{1+z}{7.5}\right)^{-1}\mbox{Mpc},
\end{eqnarray}
where ${\dot N}_{\rm \gamma,5}$ is the production rate of ionizing photons
during the luminous quasar phase for a black hole of mass $M_{\rm
bh}=10^5M_\odot$, and $f_{\rm dyn}$ is the fraction of the halo dynamical
time during which the quasar shines at its peak luminosity.

There is ample observational data on the number counts and clustering of
quasars.  We therefore compute the size of quasar HII regions using a
simple, physically-motivated, and highly successful model for high redshift
quasars that has been developed in earlier papers (Wyithe \& Loeb
2003a,2004).  The slow evolution of the quasar clustering correlation
length with redshift (Croom et al.~2001) implies that the black-hole mass
scales with halo velocity dispersion independent of redshift. This
conclusion is supported by direct observations of the $M_{\rm
bh}$--$\sigma$ relation (Shields et al. 2003).  The observed scaling and
its redshift dependence [$M_{\rm bh}\propto \sigma^5\propto M_{\rm
halo}^{5/3}(1+z)^{3/2}$] follow naturally from a scenario where black-hole
growth is limited by feedback over the galaxy dynamical time (Silk \& Rees
1998; Wyithe \& Loeb~2003a).  By associating quasar activity with major
mergers, it has been shown (Wyithe \& Loeb~2003a) that when the quasar
lifetime equals the galaxy dynamical time the observed number counts of
quasars are accurately reproduced over a wide range of redshifts, $1.5\lsim
z\lsim6.5$. The size of the quasar generated HII region ($D_{\rm q}$) may
therefore be evaluated using equation~(\ref{Dquasar}), with a black hole
mass that is related to the halo velocity dispersion (Wyithe \& Loeb~2003a)
through $M_{\rm bh} = 10^5 \left({\sigma}/{54~{\rm
km~s^{-1}}}\right)^5M_\odot$, and a lifetime that is governed by a value of
$f_{\rm dyn}=0.035$ corresponding to the typical fraction of the virial
radius occupied by the luminous component of the galaxy.  We point out that
while the association of a total quasar luminosity with a dark-matter halo
must be made via a theoretical model, the model used here correctly
predicts the number density of high redshift quasars over a wide range of
luminosities and redshifts.  The total number of ionizing photons produced
by a quasar (with a universal spectrum) during its lifetime, reflects a
fixed fraction of its black-hole rest-mass energy.

The ratio between the number of ionizing photons produced by the central
quasar and by stars is
\begin{eqnarray}
\label{Fgam}
\nonumber
\Gamma_{\rm halo}&=&1.4\left(\frac{\sigma}{100\mbox{km/s}}\right)^{2}\left(\frac{f_{\rm
dyn}}{0.035}\right)\left(\frac{f_{\star}f_{\rm
esc}}{0.003}\right)^{-1}\left(\frac{N_{\star}}{4000}\right)^{-1}\\
&\times&\left(\frac{{\dot N}_{\rm
\gamma,5}}{3\times10^{53}\mbox{s}^{-1}}\right)\left(\frac{1+z}{7.5}\right)^{3/2} .
\end{eqnarray}
Since the dependence of $\Gamma_{\rm halo}$ on halo mass is steeper than
linear for $f_{\star}=const$, quasars become increasingly more important
for higher mass galaxies.  Given that both massive stars and
quasars produce ionizing photons in short-lived episodes, the value of
$\Gamma_{\rm halo}$ reflects their net yield of ionizing photons. The HII
region generated by quasars remains as a fossil after the quasar activity
ends, since the recombination time is longer than the Hubble time at the
mean IGM density for $z\lsim 8$.  Hence, relic HII regions from prior
quasar activity must be considered when estimating the transmission of
Ly$\alpha$ photons from high redshift galaxies. In what follows we show
that this remains true even when considering the enhanced recombination
rate of the overdense IGM around massive galaxies.

\section{Str\"omgren Spheres at High-Redshift} 
\label{RHII}
Prior to reionization, galaxies generate their own Str\"omgren (HII)
spheres with characteristic sizes given by equations~(\ref{Dstar}) and
(\ref{Dquasar}).  These spheres expand into a partially neutral IGM.
If the Str\"omgren sphere is produced by a massive galaxy, then the
external neutral fraction reflects the mean volume filling factor of
all small HII regions produced by its neighboring low-mass galaxies.
The clustering of galaxies around a bright galaxy could lead to
overlap of their individual HII regions, even if the rest of the IGM
is predominantly neutral. As a result the HII region surrounding an
observed galaxy could be larger than expected from
equations~(\ref{Dstar}) and (\ref{Dquasar}) due to a cluster of faint
galaxies that are below the detection threshold of the observations.
The physical size $R_{\rm s}$ of the HII region surrounding a galaxy
of mass $M_{\rm halo}$ can be determined self-consistently with the
mean neutral fraction into which the HII region expands, through the
relation
\begin{equation}
R_{\rm s}^3 = D_{\rm halo}^3+\int_{R_{\rm vir}}^{R_{\rm s}}4\pi R^2
dR \int_{M_{\rm min}}^{M_{\rm halo}}dM \frac{D_{\rm M}^3}{\Delta_{\rm
R}^2}\frac{dn\left(R,M_{\rm halo}\right)}{dM},
\label{Rs}
\end{equation}  
where the sizes of the Str\"omgren spheres around the central halo of
mass $M_{\rm halo}$ and a galaxy of mass $M$, respectively, are
$D_{\rm halo}^3=D_{\rm halo,q}^3+D_{\rm halo,\star}^3$ and $D_{\rm
  M}^3=D_{\rm M,q}^3+D_{\rm M,\star}^3$.  Equation~(\ref{Rs}) accounts
for the higher IGM density due to infall around a massive halo through
the nonlinear density contrast $\Delta_{\rm R}(R_{\rm s},M_{\rm
  halo})=\rho/{\bar \rho}$ (where ${\bar \rho}$ is the mean background
density) at a radius $R$ from a halo of mass $M_{\rm halo}$ (Barkana~2004). The associated density profile reproduces the appropriate
two-point correlation function in the linear regime (Barkana~2004) and is
evolved into the nonlinear regime in spherical shells. The number of
galaxies surrounding the central halo is found from the
Press-Schechter~(1974) mass function of halos, which in the
vicinity of a massive halo must be modified relative to the background
universe. Within the spherical infall model, shells with density
$\rho$ at radius $R$ behave as if they are embedded in a closed
universe with a matter density parameter $\Omega_{\rm
  m}(R)=\rho/\rho_c$, where $\rho_{\rm c}=3(\dot{R}/R)^2/(8\pi G)$ and
$\dot{R}$ is the physical radial velocity of the shell (the
cosmological constant is negligible at the high redshifts of
interest.)  Thus, $dn(R,M_{\rm halo})/dM$ is the physical number
density of halos per unit mass computed in a universe with a density
parameter $\Omega_{\rm m}(R)$ (with a normalization and a growth
factor based on the local shell density). Each HII region reaches its
final comoving volume when the production rate of ionizing photons is
balanced by the recombination rate interior to that volume. In this
phase, the HII volume is inversely proportional to the square of the
mean density times a clumping factor.  The clumping is accounted for
by $N_{\rm reion}$, but an additional factor of $\Delta_{\rm R}^{-2}$
needs to be included on the right-hand-side of equation~(\ref{Rs}) to
account for recombinations at the mean density within each shell.  We
assume that the most massive galaxy interior to the HII region is the
one observed.  This conservative assumption is reflected in the upper
limit of the mass integration. The value $M_{\rm min}$ is the minimum
mass (Barkana \& Loeb~2001), below which gas infall is suppressed from an ionized
region that has been heated to $10^{4}$K.

\begin{figure*}[t]
\epsscale{1.5}  \plotone{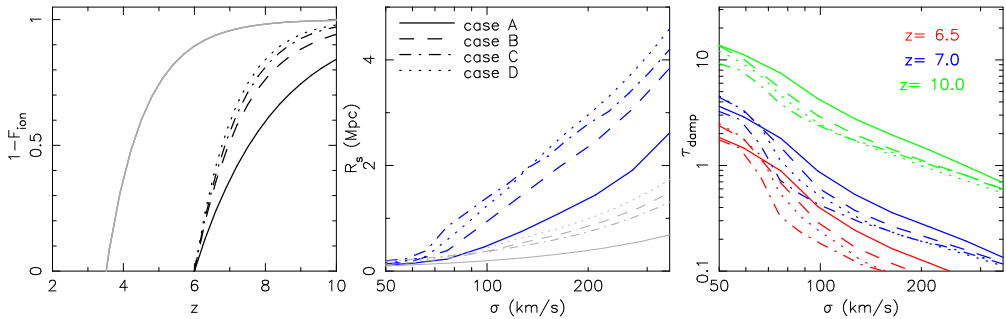}
\caption{\label{fig1} The physical size of cosmological HII regions, and
the detectability of high redshift galaxies. {\it Left}: The neutral
fraction of the intergalactic hydrogen vs. redshift [see
equation~(\ref{FH})], with the constraints that hydrogen is reionized at
$z=6$ and that helium is doubly-reionized at $z=3.5$.  The gray, left-most
curve shows the neutral fraction of HeII.  {\it Center}: The size of an HII
region at ($z=7$) due to clustered faint sources around a central galaxy
with a velocity dispersion $\sigma$. Also shown (gray curves) are values
for the naive estimate of the size of the HII region, assuming a fully
neutral IGM at mean density [Eq.~(\ref{Dstar})].  {\it Right}: Optical
depth for Ly$\alpha$ absorption by the damping wing of the distant IGM, at
the center of an HII region around a galaxy with a velocity dispersion
$\sigma$. From left to right, the different sets of lines correspond to
redshifts $z=6.5$, $z=7$ and $z=10$, respectively. The different line types
mark the four cases {\bf A-D} (see main text) labeled in the middle panel.
}
\end{figure*}

Next we specify the free parameters in our model. The absorption spectra of
the highest redshift quasars indicate that the most recent reionization of
hydrogen occurred at $z\sim6$ (Fan et al.~2001; Wyithe \& Loeb~2004).  Moreover, there is evidence for
a temperature rise in the Ly$\alpha$ forest associated with the
reionization of HeII at $z\sim3.5$ (Theuns et al.~2002).  We set the free parameters
by requiring that hydrogen be reionized by $z=6$, and helium be doubly
reionized by $z=3.5$.  Population-II stars do not have spectra that are
sufficiently hard to ionize HeII, and so in the absence of Population-III
stars at low redshifts the ionization fraction of HeII is determined by
quasars. The global ionization fractions of intergalactic hydrogen and
helium are therefore
\begin{eqnarray}
\label{FH}
\nonumber
F_{\rm ion,HII}&=&\int_{M_{\rm
min}}^{\infty}dM\frac{dn}{dM}\frac{4\pi}{3}\left(D_{\rm M,q}^3+D_{\rm
M,\star}^3\right)\hspace{3mm}\mbox{and}\\
F_{\rm ion,HeIII}&=&\int_{M_{\rm min}}^{\infty}dM\frac{dn}{dM}\frac{4\pi}{3} D_{\rm
M,He,q}^3.
\end{eqnarray}
where $D_{\rm M,He,q}\approx 0.85D_{\rm M,q}$ assuming the median spectral
energy distribution of quasars (Telfer et al.~2002), and that $\sim 50\%$ of the
photons of sufficient energy to double-ionize helium are utilized in the
reionization of hydrogen (Wyithe \& Loeb~2003b).  The ratio ${\dot N}_{\rm
\gamma,5}/N_{\rm reion}$ may be determined from the requirement that
$F_{\rm ion,HeIII}$ approach unity at $z\sim3.5$. Assuming, for simplicity,
that $N_{\rm reion}=const$ and that the spectral slope of quasars blueward
of the ionization threshold of hydrogen is specified by their median
spectral template (Telfer et al.~2002), the ratio $f_{\star}f_{\rm esc}/N_{\rm
reion}$ can then be determined from the requirement that $F_{\rm ion,HII}$
approach unity at $z\sim6$. Gas clumping is expected to increase with
cosmic time.  By calibrating $N_{\rm reion}$ based on the reionization of
helium and hydrogen at $z\sim3.5$ and $6$, we therefore overestimate the
significance of recombinations at earlier times and derive conservative
results for the minimum transmission of the Ly$\alpha$ line. We note that
even though our model is calibrated by requiring that hydrogen reionization
be completed at $z\sim6$ by Pop-II stars and quasars, an early
generation of massive Pop-III stars may be required to produce a first
partial or full reionization of hydrogen at $z\sim15$--$20$. An early 
reionization would in provide consistency between the quasar absorption at
$z\sim6$ and the large scale polarization anisotropies of the
microwave background detected by the {\it Wilkinson Microwave Anisotropy
Probe} ({\it WMAP}; Wyithe \& Loeb~2003b; Cen~2003).

Reionization histories may now be calculated based on
equation~(\ref{FH}) for different star formation prescriptions, with
${\dot N}_{\rm \gamma,5}$ and $f_{\star}$ adjusted to meet the above
global requirements. We adopt feedback-regulated prescriptions (Dekel \& Woo~2002; Wyithe \& Loeb~2003a) in which the star formation efficiency in a galaxy depends
on its velocity dispersion $\sigma$, with $f_{\star}=0$ for
$\sigma<\sigma_{\rm min}$, $f_{\star}=f_{\rm \star,crit}
\left({\sigma}/{\sigma_{\rm crit}}\right)^2$ for $\sigma_{\rm
  min}<\sigma<\sigma_{\rm crit}$, and $f_{\star}=f_{\rm \star,crit}$ for
$\sigma>\sigma_{\rm crit}$.  We consider four cases in this paper:
({\bf A})~$\sigma_{\rm min}=\sigma_{\rm crit}=10~{\rm km~s^{-1}}$;
({\bf B})~$\sigma_{\rm min}=10~{\rm km~s^{-1}}$, $\sigma_{\rm
  crit}=120~{\rm km~s^{-1}}$; ({\bf C})~$\sigma_{\rm min}=\sigma_{\rm
  crit}=50~{\rm km~s^{-1}}$; and ({\bf D})~$\sigma_{\rm min}=50~{\rm
  km~s^{-1}}$, $\sigma_{\rm crit}=120~{\rm km~s^{-1}}$. The values of
$\sigma_{\rm min}=50~{\rm km~s^{-1}}$, and $\sigma_{\rm min}=10~{\rm
  km~s^{-1}}$, correspond to the threshold for infall of gas from a
photo-ionized IGM and the threshold for atomic hydrogen cooling,
respectively.  The value of $\sigma_{\rm crit}=120~{\rm km~s^{-1}}$ is
calibrated empirically based on the local sample of galaxies (Kauffman et al.~2003; Dekel \& Woo 2002).  We also examine the case of $\sigma_{\rm crit}=\sigma_{\rm
  min}$ to reflect our ignorance about whether a similar feedback
operates at high redshifts, when the dynamical time within galaxies
becomes comparable to the lifetime of massive stars. We have adopted
the set of cosmological parameters determined by {\it WMAP} (Spergel et al. 2003), and find that a constraint of late reionization at $z\sim6$
requires values for the parameter combination
$N_{\star}f_{\rm \star,crit}f_{\rm esc}/N_{\rm reion}$ of 11, 2, 22, and 17 for
models {\bf A-D}.

In the central panel of Figure~\ref{fig1}, we plot the Str\"omgren
radius $R_{\rm s}$ as a function of the velocity dispersion $\sigma$
of a central galaxy at a redshift of $z=7$.  The four curves
correspond to cases {\bf A-D}. These calculations of $R_{\rm s}$ are
self-consistent with the average neutral fraction of the IGM, which is
obtained from the reionization histories (Eq.~\ref{FH}), and
plotted in the left panel of Figure~\ref{fig1}. A rapid, stellar
reionization that completes by $z=6$ would imply a neutral fraction of
$x_{\rm HI}\sim0.2$--$0.5$ at $z\sim6.5-7$, consistent with the
analysis of the HII regions around the highest redshift quasars
(Wyithe \& Loeb~2004). At $z=7$, the HII regions approach a radius of
$R_{\rm s}\sim 0.5-1$~Mpc for $\sigma\sim100~{\rm km~s^{-1}}$,
comparable to the values inferred for recently discovered galaxies
(Kneib et al.~2004; Rhoads et al.~2004). Galaxies with
$\sigma\sim300~{\rm km~s^{-1}}$, corresponding to the hosts of the
luminous $z\gsim6$ quasars, have $R_{\rm s}\sim3-4$Mpc in agreement
with the observed sizes of their Str\"omgren spheres (White et
al.~2003).  For comparison, we also show by gray lines the naive
estimate for the size of the HII region using only the stars within
the central halo, and assuming a fully neutral IGM at the mean density
(Eq.~(\ref{Dstar}).

\begin{figure*}[t]
\epsscale{1.5}  \plotone{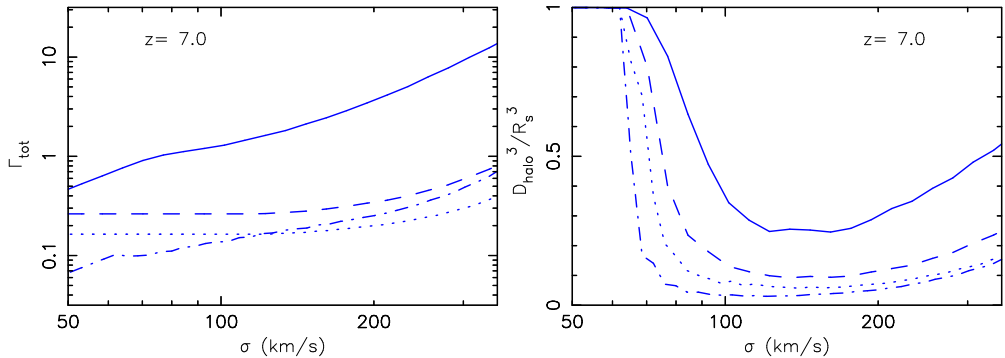}
\caption{\label{fig2} Comparisons between the contributions to the HII
region of stars relative to quasars, and of the central galaxy relative to
the surrounding cluster of galaxies. {\it Left}: The ratio between the
number of ionizing photons contributed to the HII region by quasars and by
stars as a function of the velocity dispersion of the host galaxy
[Eq.~(\ref{Gamtot})]. {\it Right}: The fraction of the HII region volume
contributed by the central galaxy as a function of the velocity dispersion
of the host galaxy.  In each panel the different line types correspond to
cases {\bf A-D} as in Figure~\ref{fig1} for galaxies at $z=7$.}
\end{figure*}

\section{Suppression of the Ly$\alpha$ Emission Line} 

At $z\sim6.5-7$, the ionized regions around galaxies with
$\sigma\gsim100~{\rm km~s^{-1}}$ are sufficiently large to allow
transmission of their Ly$\alpha$ emission line despite the red damping wing
of Ly$\alpha$ absorption from the surrounding IGM. It is therefore not
surprising that these galaxies are observed, even if they reside in an IGM
that is significantly neutral globally.  The optical depth $\tau_{\rm
damp}$ to Ly$\alpha$ absorption at the galaxy redshift
(Miralda-Escude~1998) is plotted as a function of $\sigma$ in the right
hand panel of Figure~\ref{fig1} for galaxies at $z\sim6.5$, $z\sim7$ and
$z\sim10$.  Here the size of the HII region is based on
equation~(\ref{Rs}), while the neutral fraction outside the HII region was
computed from equation~(\ref{FH}).  At $z\sim10$, a galaxy must possess
$\sigma\gsim200-300~{\rm km~s^{-1}}$ in order to have $\tau_{\rm
damp}<1$. However at $z=6.5-7$, $\tau_{\rm damp}\lsim1$ for galaxies with
$\sigma\gsim70-100~{\rm km~s^{-1}}$. Hence, observations of galaxies at
$z\sim6.5$--7 (Kodaira et al., 2003; Rhoads et al.~2004; Kurk et al. 2004;
Stern et al. 2004) are fully consistent with reionization completing at
$z\sim6$. The requirement that HII and HeIII be reionized at $z\sim6$ and
$z\sim3.5$ respectively, make this conclusion robust against variations in
$f_{\rm \star,crit}$ or $N_{\rm reion}$.

\section{Contributions to the Ionizing Flux} 

We now explicitly examine the individual contributions from quasars
and from the clustering of faint sources to the HII region around
massive galaxies. First, we find the ratio between the quasar and
stellar fluxes within the HII region, $\Gamma_{\rm tot}$.  The flux
ratio $\Gamma_{\rm halo}$ from the central halo is obtained from
equation~(\ref{Fgam}). The mean flux ratio $\langle \Gamma_{\rm
  gal}\rangle$ from other galaxies within the HII region of radius
$R_{\rm s}$ is then
\begin{eqnarray}
\nonumber
\label{Fgamgal}
\nonumber
\langle \Gamma_{\rm gal}\rangle &=& \left(R_{\rm s}^{3}-D_{\rm
halo}^3\right)^{-1}\\
 &&\hspace{-10mm}\times\int_{R_{\rm vir}}^{R_{\rm s}}4\pi R^2dR\int_{M_{\rm
min}}^{M_{\rm halo}} dM \frac{D_{\rm M}^3}{\Delta_{\rm R}^2} 
\frac{dn(M_{\rm halo},R)}{dM} \Gamma,
\end{eqnarray}
and the average flux ratio within the HII region is therefore
\begin{equation}
\label{Gamtot}
\Gamma_{\rm tot}=\frac{\left(R_{\rm s}^3-D_{\rm
halo}^3\right)\langle \Gamma_{\rm gal}\rangle + D_{\rm halo}^3\Gamma_{\rm
halo}}{R_{\rm s}^{3}}.
\end{equation}
The left-hand-panel of Figure~\ref{fig2} shows $\Gamma_{\rm tot}(\sigma)$
for cases {\bf A-D} at $z=7$. Quasars make a substantial contribution to
the Str\"omgren volume around massive galaxies, particularly in case-{\bf A}. 
However the cluster of smaller galaxies 
surrounding the central halo contributes an overall
ionizing luminosity that is mostly stellar, resulting in a value of 
$\Gamma_{\rm tot}$ that is smaller than $\Gamma_{\rm halo}$.

The fraction of the HII region excavated by the central galaxy is found
from the ratio $D_{\rm halo}^3/R_{\rm s}^3$. This quantity is shown in the
right panel of Figure~\ref{fig2} as a function of $\sigma$ for galaxies at
$z=7$.  The regions around massive galaxies are predominantly ionized by
the surrounding cluster of smaller galaxies.  The luminous quasars
discovered at $z\gsim6$ have, during their active phase, luminosities in
excess of the time-averaged luminosity of these clusters, and are therefore
observed as the source of their HII region. 

\section{Summary}
The massive galaxies discovered prior to reionization should have
Str\"omgren spheres that are much larger than naively expected from the UV
continuum flux of their stars. This increase arises for two reasons: {\it
(i)} there is an additional contribution from earlier quasar activity
within the central and surrounding galaxies, and {\it (ii)} there is a
biased clustering of faint sources surrounding the central galaxy which
results in overlap of nearby HII regions at a time before overlap is
achieved throughout the universe. The large extent of the HII regions allow
transmission of Ly$\alpha$ photons from galaxies. We conclude that the
discovery of Ly$\alpha$ emission from bright galaxies at $z>6.5$ does not
indicate a highly ionized IGM. Nevertheless, the strength of the Ly$\alpha$
damping increases as the galaxy luminosity decreases or the redshift
increases (see the right-hand panel of Fig. \ref{fig1}). A neutral IGM
would therefore suppress the low-luminosity end of the Ly$\alpha$
luminosity function of galaxies at high redshifts.

\acknowledgements 
We thank Daniel Stern for comments on the manuscript.
This work
was supported in part by NASA grant NAG 5-13292, and by NSF grants
AST-0071019, AST-0204514 for AL.

\end{document}